\documentclass[preprint,prd,showpacs,superscriptaddress,preprintnumbers]{revtex4}

\usepackage{amsmath}
\usepackage{amssymb}
\usepackage{bm}
\usepackage{dcolumn}
\usepackage{graphicx}
\usepackage{subfigure}

\newenvironment{resulttable}{%
\begin{ruledtabular}
\begin{tabular}{lc@{\hskip-3em}d@{\hskip2em}c@{\hskip1em}c}
\makebox[7em][l]{Integral} & $n_F$ & \makebox[-5em]{Value (Error)}   & Sampling per & No. of \\ [-.5ex]  
                           &       & \makebox[-5em]{including $n_F$} & iteration    & iterations \\ 
\hline

}{%
\end{tabular}
\end{ruledtabular}
}

\newenvironment{renomtable}{%
\begin{ruledtabular}
\begin{tabular}{l@{\hskip-3em}d@{\hskip4em}l@{\hskip-3em}d}
Integral & \makebox[-4em]{Value(Error)} & 
Integral & \makebox[-4em]{Value(Error)} \\
\hline
}{%

\end{tabular}
\end{ruledtabular}
}

\begin{document}

\date{\today}

\preprint{RIKEN-QHP-1}

\title{%
Tenth-Order Lepton Anomalous Magnetic Moment --
Sixth-Order Vertices Containing Vacuum-Polarization Subdiagrams
}


\author{Tatsumi Aoyama}
\affiliation{Kobayashi-Maskawa Institute for the Origin of Particles and the Universe (KMI), Nagoya University, Nagoya, 464-8602, Japan}
\affiliation{Nishina Center, RIKEN, Wako, Japan 351-0198 }

\author{Masashi Hayakawa}
\affiliation{Department of Physics, Nagoya University, Nagoya, Japan 464-8602 }
\affiliation{Nishina Center, RIKEN, Wako, Japan 351-0198 }

\author{Toichiro Kinoshita}
\affiliation{Laboratory for Elementary-Particle Physics, Cornell University, Ithaca, New York, 14853, U.S.A }
\affiliation{Nishina Center, RIKEN, Wako, Japan 351-0198 }

\author{Makiko Nio}
\affiliation{Nishina Center, RIKEN, Wako, Japan 351-0198 }

\begin{abstract}
This paper reports the values of
contributions to the electron $g\!-\!2$ from
300 Feynman diagrams of the gauge-invariant Set III(a) 
and 450 Feynman diagrams of the gauge-invariant  Set III(b).
The evaluation is carried out in two versions.  {\it Version A} is
to start from the sixth-order magnetic anomaly $M_6$
obtained in the previous work.
The mass-independent contributions of Set III(a) and Set III(b)
are $2.1275~(2)$ and $3.3271~(6)$ in units of $(\alpha/\pi)^5$, respectively.
{\it Version B} is based on the recently-developed
automatic code generation scheme.
This method yields $2.1271~(3)$ and $3.3271~(8)$
in units of $(\alpha/\pi)^5$, respectively.
They are in excellent agreement with the results of the first method 
within the uncertainties of numerical integration.
Combining these results as statistically independent
we obtain the best values, $2.1273~(2)$, and $3.3271~(5)$
times $(\alpha/\pi)^5$, for the mass-independent contributions
of the Set III(a) and Set III(b), respectively.
We have also evaluated mass-dependent contributions
of diagrams containing muon and/or tau-particle loop.
Including them the total contribution of Set III(a) is $2.1349~(2)$ 
and that of Set III(b) is $3.3299~(5)$ in units of  $(\alpha/\pi)^5$.
The total contributions to the muon $g\!-\!2$ of various 
leptonic vacuum-polarization loops of Set III(a) and Set III(b) 
are $112.418~(32)$ and $15.407~(5)$
in units of  $(\alpha/\pi)^5$, respectively.
\end{abstract}

%

\maketitle

\section{Introduction}
\label{sec:intro}

The anomalous magnetic moment $g\!-\!2$ of the electron has played 
the central role in testing the validity of quantum electrodynamics (QED)
as well as the Standard Model.
The latest measurement of $a_e \equiv (g-2)/2$ by the Harvard group has reached the precision
of $0.24 \times 10^{-9}$ 
\cite{Hanneke:2008tm,Hanneke:2010au}:
\begin{eqnarray}
a_e(\text{HV08})= 1~159~652~180.73~ (0.28) \times 10^{-12} ~~~[0.24 \text{ppb}]
~.
\label{a_eHV08}
\end{eqnarray}
At present the theoretical prediction consists of 
QED corrections of up to the eighth order
\cite{Kinoshita:2005sm,Aoyama:2007dv,Aoyama:2007mn},
and hadronic corrections \cite{Davier:2010nc,Teubner:2010ah,Krause:1996rf,
Melnikov:2003xd,Bijnens:2007pz,Prades:2009tw,Nyffeler:2009tw} 
and electro-weak corrections \cite{Czarnecki:1995sz,Knecht:2002hr,Czarnecki:2002nt} scaled down from their contributions to the muon $g\!-\!2$. 
To compare the theory with the measurement 
(\ref{a_eHV08}),
we also need  the value of the fine structure constant $\alpha$
determined by a method independent of $g\!-2\!$ .
The best value of such an $\alpha$ has been obtained recently
from the measurement of $h/m_{\text{Rb}}$, the ratio of the Planck constant
and the mass of Rb atom,  
combined with the very precisely known Rydberg constant 
and $m_\text{Rb}/m_e$: \cite{Bouchendira:2010es}
\begin{eqnarray}
\alpha^{-1} (\text{Rb10}) = 137.035~999~037~(91)~~~[0.66 \text{ppb}].
\label{alinvRb10}
\end{eqnarray}  
With this  $\alpha$ the theoretical prediction of $a_e$ becomes 
\begin{eqnarray}
a_e(\text{theory}) = 1~159~652~181.13~(0.11)(0.37)(0.77) \times 10^{-12},
\label{a_etheory}
\end{eqnarray}
where the first, second, and third uncertainties come
from the calculated eighth-order QED term, the tenth-order estimate, and the
fine structure constant (\ref{alinvRb10}), respectively.
The theory (\ref{a_etheory})
is thus in good agreement with the
experiment (\ref{a_eHV08}):  
\begin{eqnarray}
a_e(\text{HV08}) - a_e(\text{theory}) = -0.40~ (0.88) \times 10^{-12},
\end{eqnarray}
proving that QED (Standard Model) is in good shape even at this very high
precision.

An alternative test of QED is to compare $\alpha$(Rb10) with
the value of $\alpha$ determined from the experiment and theory of $g\!-2\!$~:  
\begin{eqnarray}
\alpha^{-1}(a_e 08) = 137.035~999~085~(12)(37)(33)~~~[0.37 \text{ppb}],
\label{alinvae}
\end{eqnarray}
where the first, second, and third uncertainties come
from the eighth-order QED term, the tenth-order estimate, and the
measurement of $a_e$(HV08), respectively.
Although the uncertainty of
$\alpha^{-1} (a_e 08)$ in (\ref{alinvae}) is a factor 2 smaller than
$\alpha$(Rb10), it is not a firm factor since it
 depends on the estimate of the tenth-order term,
which is only a crude guess \cite{Mohr:2008fa}.
For a more stringent test of QED, it is obviously necessary to calculate
the actual value of the tenth-order term.
In anticipation of  this challenge we launched 
a systematic program 
several years ago
to evaluate the complete tenth-order term
\cite{Kinoshita:2004wi,Aoyama:2005kf,Aoyama:2007bs}.


The 10th-order QED contribution to the 
anomalous magnetic moment of an electron can be written as
\begin{equation}
	a_e^{(10)} 
	= \left ( \frac{\alpha}{\pi} \right )^5 
          \left [ A_1^{(10)}
	+ A_2^{(10)} (m_e/m_\mu) 
	+ A_2^{(10)} (m_e/m_\tau) 
	+ A_3^{(10)} (m_e/m_\mu, m_e/m_\tau) \right ],
\label{eq:ae10th}
\end{equation}
where $m_e/m_\mu = 4.836~331~71~(12) \times 10^{-3}$
and $m_e/m_\tau = 2.875~64~(47) \times 10^{-4}$ \cite{Mohr:2008fa}.
In the rest of this article the factor
$ \left ( \frac{\alpha}{\pi} \right )^5 $ 
will be suppressed for simplicity.

The contribution to the mass-independent term $A_1^{(10)}$ can be
classified into six gauge-invariant sets, further divided into
32 gauge-invariant subsets depending on the nature of closed
lepton loop subdiagrams.
Thus far, numerical results of 27 gauge-invariant subsets, 
which consist of 3106 vertex diagrams, 
have been published \cite{Kinoshita:2004wi,Aoyama:2008gy,Aoyama:2008hz,Aoyama:2010yt,Aoyama:2010pk,Aoyama:2010zp,Aoyama:2011rm}.
Five of these 27 subsets were also calculated 
analytically \cite{Laporta:1994md,Aguilar:2008qj}.
Our calculation is in good agreement with the analytic results.

\begin{figure}[ht]
\includegraphics[width=13cm]{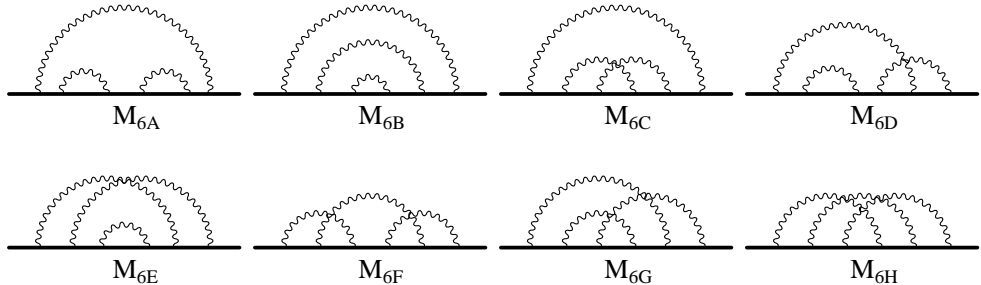}
\caption{
\label{fig:M6} 
The sixth-order $q$-type diagrams. The solid line represents the electron in a constant
magnetic field. The time reversal diagrams of $M_{6D}$ and $M_{6G}$ are
omitted for simplicity. 
}
\end{figure}

In this paper we report the evaluation of the tenth-order lepton  $g\!-\!2$
from two gauge-invariant subsets called Set III(a) and Set III(b). 
These diagrams are built from 
the magnetic moment contribution $M_6$ 
(shown in Fig. ~\ref{fig:M6})
which consists of 50 proper sixth-order vertices of
three-photon-exchange type,
namely diagrams without closed lepton loops (and called $q$-type. See
Ref.~\cite{Aoyama:2007dv}  for the definition of $q$-type.),
by insertion of various lepton vacuum-polarization loops.  

\begin{figure}
\includegraphics[scale=0.8]{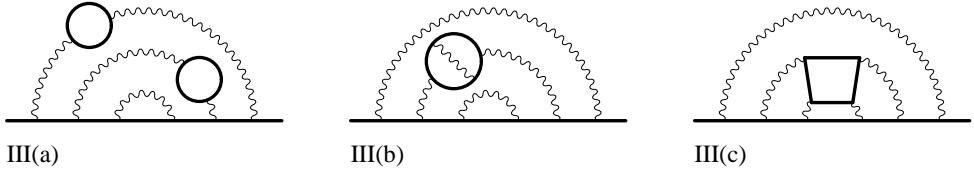}
\caption{Typical diagrams of Set III.}
\label{fig:set3}
\end{figure}

\begin{description}
\item[Set III(a).] 
Diagrams obtained by inserting two 
second-order vacuum-polarization function $\Pi_2$'s in $M_6$. 
The number of vertex diagrams contributing to $A_1^{(10)}$ is 300. 

\item[Set III(b).] 
Diagrams obtained by inserting 
the fourth-order vacuum-polarization function $\Pi_4$ in $M_6$, where 
$\Pi_4$ is the sum of three fourth-order vacuum polarization loops. 
The number of vertex diagrams contributing to $A_1^{(10)}$ is 450. 
\end{description}

Another set ( Set III(c) of Fig. \ref{fig:set3})
consists of diagrams obtained by inserting a light-by-light scattering
subdiagram $\Lambda_4$ in $M_6$.
The total number of these diagrams contributing to $A_1^{(10)}$ is 390. 
Since it has a structure different from those of Sets III(a) and III(b),
it will be treated  in a separate paper.

Evaluation of Set III(a) and Set III(b)
is carried out in two ways.  {\it Version A} is
to start from the {\sc FORTRAN} code of the
sixth-order anomalous magnetic moment $M_6$,  which was obtained in
previous works \cite{Cvitanovic:1974uf} and
known to give the result identical with the analytic result \cite{Laporta:1996mq}. 
It is thus easy to establish the validity of these 
{\sc FORTRAN} codes for Sets III(a) and III(b).  

We also evaluate these sets 
by an alternative method, {\it Version B},  using {\sc FORTRAN} codes generated from scratch by
the recently developed automatic code generation scheme \cite{Aoyama:2007dv,Aoyama:2005kf}.
This approach deals with the
UV renormalization as well as IR subtraction terms as integral parts of automation.
In carrying out this automation scheme, we found it useful to
construct IR subtraction terms in a 
different manner from that of {\it Version A} \cite{Aoyama:2005kf}.
Thus, {\it Version B}  provides  an independent confirmation 
of {\it Version A}.  At the same time it helps to verify
the automated code generation scheme, which is developed primarily to deal
with the vastly more difficult problem of Set V,
which consists of 6354 vertex diagrams with pure radiative correction.

As is well-known, the insertion of vacuum-polarization loop
such as $\Pi^{(2)}$ and $\Pi^{(4)}$  in
an internal photon line of momentum $q$ can be expressed 
as a superposition of massive vector propagators
\begin{equation}
	\int_{4m^2}^\infty \frac{d\sigma \rho(\sigma)}{q^2-\sigma},
\label{eq:superposition}
\end{equation}
where $m$ is the mass of the lepton forming the closed loop and
$\sigma$ is the square of mass of the vector particle
and $\rho$ is the spectral function.
This enables us to obtain Feynman-parametric integrals for Set III(b)
by simply replacing the relevant photon mass squared by $\sigma$
and integrating over $\sigma$.
It can also be applied to diagrams of Set III(a) which
contain two vacuum-polarization loops in {\it different} photon lines.
This subset of Set III(a) will be denoted as Set III(a$_d$) henceforth.

\begin{figure}[t]
\includegraphics[scale=1.0]{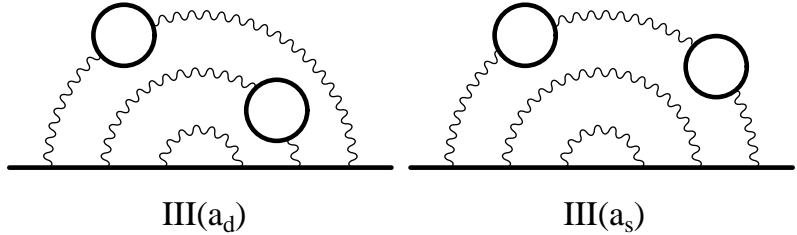}
\caption{\label{fig:set3a}
Typical tenth-order diagrams of Set III(a) obtained by insertion of 
two second-order vacuum-polarization loops $\Pi_2$ in lepton 
diagrams of the three-photon-exchange type.  
The subset III(a$_d$) consists of diagrams in  which $\Pi_2$
are inserted in different photon lines, while
the subset III(a$_s$) consists of diagrams in  which $\Pi_2$
are inserted in the same photon line.
There are 
150 diagrams in each subset.}
\end{figure}

The Set III(a) also contains diagrams in which 
two vacuum-polarization loops are inserted in the {\it same} photon line,
which will be denoted as Set III(${\rm a}_s$).  For these diagrams
a slight extension of Eq.~(\ref{eq:superposition}) is required.
When two vacuum polarization loops are inserted in a photon
line of momentum $q$, the result,
omitting integrations for simplicity,
is given by the left-hand-side of the following equation,
which can be rewritten in the form on the right-hand-side:
\begin{equation}
	\frac{1}{q^2-\sigma_a} q^2 \frac{1}{q^2 - \sigma_b} 
	\equiv
	\frac{\sigma_a}{\sigma_a -\sigma_b} \frac{1}{q^2- \sigma_a}
	-\frac{\sigma_b}{\sigma_a -\sigma_b} \frac{1}{q^2- \sigma_b}.
\label{eq:superposition2}
\end{equation}
Note that the right-hand-side is a linear combination of propagators
of mass-square $\sigma_a$ and $\sigma_b$
with coefficients $\sigma_a/(\sigma_a -\sigma_b)$ 
and $-\sigma_b/(\sigma_a -\sigma_b)$ .
This enables us to write the Feynman-parametric integrals
for the diagrams in Set III(a$_s$) 
by a simple extension of $M_6$ integrals.
Eq.~(\ref{eq:superposition2}) can be readily extended to the case in which
 three or more vacuum-polarization loops are inserted in the same photon line.

These adaptations require a slight modification of the numerator
function $V$, which, for $M_6$, is given by
\begin{align}
V_0 &= \sum_{i=1}^{5} z_i( 1- A_i)m_e^2, \nonumber \\   
V   &= V_0 + (z_a + z_b + z_c) \lambda ^2~,
\label{defv}
\end{align}
where $z_i~(i=1,\cdots,5)$, and $z_j~(j=a,b,c)$  are Feynman parameters 
assigned to the fermion propagators
and the photon propagators, respectively. $m_e$ and $\lambda$ are masses of the electron and photon, respectively.
$A_i~(i=1,\cdots,5)$ are {\it scalar currents} flowing 
in the fermion line $i$ (see the exact definition of $A_i$ in Ref.~\cite{Cvitanovic:1974uf}
). 
$A_i$ is expressed by the Feynman parameters and its expression
depends on the structure of a  diagram. But, the expression of $V$
in terms of $A_i$ is identical for all diagrams of $M_6$.

When one vacuum-polarization function is inserted in a photon line, 
we must replace the mass square $\lambda^2$ of the photon in Eq.~(\ref{defv})
by $p(t)$:
\begin{equation}
\lambda^2 \longrightarrow  p(t) \equiv  \frac{4 m_{vp}^2}{1 - t^2},
\label{loopmass}
\end{equation}
where $m_{vp}$ is the rest mass of the fermion forming the vacuum-polarization 
loop and  the interval $4m_{vp}^2 \leq \sigma < \infty$ of Eq.~(\ref{eq:superposition})
is mapped onto $(0 \leq t < 1)$ for the sake of convenience.

When two vacuum-polarization functions are inserted in the same 
photon line $a$, 
it follows from Eq.~(\ref{eq:superposition2}) that the denominators
must be modified as follows:
\begin{align}
&\frac{1}{V} \longrightarrow \frac{V_0}{V_1 V_2} , \nonumber \\
&\frac{1}{V^2} \longrightarrow \frac{V_0^2 -z_a^2 p_1(t_1) p_2(t_2)}{(V_1 V_2 )^2 } , \nonumber \\
&\frac{1}{V^3} \longrightarrow \frac{V_0^3 
-3 V_0 z_a^2 p_1(t_1) p_2(t_2) 
- z_a^3 p_1(t_1) p_2(t_2) (p_1(t_1)+ p_2(t_2)) }{( V_1 V_2)^3 }~,
\end{align}
where  
\begin{equation}
V_i \equiv V_0 + z_a p_i (t_i)+ (z_b + z_c) \lambda^2, ~~~ i=1,2,~
\end{equation}
for the first or second vacuum-polarization functions.

Throughout this article we use the exact renormalized forms 
of $\Pi_2$ and $\Pi_4$ instead of intermediately renormalized forms to
take advantage of the known analytic forms of their spectral functions \cite{Kallen:1955fb}.


\section{Set III(\text{a}) }
\label{subsec:set3a}

Diagrams belonging to the Set III(a) are generated 
by inserting two second-order
vacuum-polarization loops  $\Pi_2$ in
the photon lines of $M_6$.
Using an identity derived from the Ward-Takahashi identity
\cite{Kinoshita:2004wi} and
time-reversal invariance and summing up all possible
insertions of the photon spectral function reduce the
number of independent integrals from 300 to 16.  
For programming purpose it is convenient to treat
 Set III(${\rm a}_d$) and  Set III(${\rm a}_s$) separately.

%


\subsection{Set III(${\rm a}_d$)}
\label{subsec:set3a_d}

Let $M_{6 \alpha ,P2P2}$ be the magnetic moment 
projection of the Set III(${\rm a}_d$) generated from a self-energy
diagram $M_{6\alpha}$ ($\alpha$ =A through H) 
by insertion of two electron vacuum-polarization loops
$\Pi_2$ in different photon lines
(see Fig.~\ref{fig:set3a}). 
The subscript $P2P2$ implies that two second-order 
vacuum polarization function $P2$'s are
inserted in different photon lines of the proper diagram $M_{6\alpha}$.
To be precise 
$M_{6 \alpha ,P2P2}$ should be written as $M_{6 \alpha ,P2P2}^{(l_1 l_2 l_3)}$,
where the first superscript $l_1$ refers to the open lepton line
and $l_2$ and $l_3$ refer to closed lepton loops.
When $l_1$,  $l_2$, and $l_3$ are identical
so that $M_{6\alpha ,P2P2}$ is mass-independent,
 we omit the superscripts for simplicity.
Distinction by superscript becomes necessary in Sec. \ref{subsubsec:massdependence_3ad}
where mass-dependent terms are treated.


\subsubsection{Electron $g\!-\!2$: Version A}
\label{subsubsec:set3a_d_electron_firstmethod}

In {\it Version A} the renormalized
contribution of the diagrams of Set III(${\rm a}_d$) can 
be written as \cite{Kinoshita:1990} 
\begin{equation}
a_e^{(10)} [\text{Set III}({\rm a}_d)] = \sum_{\alpha = A}^H  a_{6 \alpha ,P2P2} ,
\label{aIIIad}
\end{equation}
with
\begin{equation}
	a_{6 \alpha ,P2P2} 
	= 
	\Delta M_{6 \alpha ,P2P2} 
	+ 
	\text{residual renormalization terms},
\end{equation}
where $\Delta M_{6 \alpha, P2P2}$ is the UV- and IR-finite
part of $M_{6 \alpha, P2P2}$ after all divergences
are removed by intermediate renormalization by $K_S$ and
$I_R$ operations.   
See Ref.~\cite{Kinoshita:1990} for definitions of $K$-operation and $I$-operation.

When summed over all diagrams of Set III(${\rm a}_d$), 
the UV- and IR-divergent pieces cancel out 
and the total contribution to $a^{(10)}$ can be written as a sum of
finite pieces \cite{Kinoshita:2004wi}:
\begin{align}
       a_e^{(10)} [\text{Set III}({\rm a}_d): \text{\it Ver.~A}] 
	& = 
	\sum_{ \alpha = A}^H \Delta M_{6 \alpha ,P2P2}
\nonumber \\
	& -3 \Delta B_{2,P2} \Delta M_{4,P2} -3 \Delta B_2  \Delta M_{4,P2P2}
\nonumber \\
        & + \Delta \delta m_{4,P2} ( M_{2^{*} ,P2} [I] - M_{2^* , P2} )
         + \Delta \delta m_{4,P2P2} (M_{ 2^{*}} [I] -M_{2^*} )  
\nonumber \\
	& - [ \Delta B_{4,P2} + 2 \Delta L_{4,P2} - 4\Delta B_2 \Delta B_{2,P2} ] M_{2,P2}
\nonumber \\
	& - [ \Delta B_{4, P2P2} + 2 \Delta L_{4,P2P2}-2 (\Delta B_{2,P2})^2 ]
M_2 ,
\label{aIIIyy}
\end{align}
where the number of vertex diagrams represented by
$\Delta M_{6 \alpha ,P2P2}$ is 15 for $\alpha = A, B, C$, $E, F, H$
and 30 for $\alpha = D, G$ (see Fig. \ref{fig:M6}). 
We use the compactified notations for the magnetic moment, mass renormalization
constant, wave-function renormalization constant, and vertex renormalization constants of fourth order \cite{Kinoshita:1990}:
\begin{align}
	\Delta M_4 &\equiv \Delta M_{4a} + \Delta M_{4b},
\nonumber \\
	\Delta \delta m_4 &\equiv \Delta \delta m_{4a} + \Delta \delta m_{4b},
\nonumber \\
	\Delta B_4 &\equiv \Delta B_{4a} + \Delta B_{4b},
\nonumber \\
	\Delta L_4 &\equiv \sum_{i=1}^3 (\Delta L_{4a,i} +  \Delta L_{4b,i}),
\label{defM4dm4B4L4}
\end{align}
where $4a$ and $4b$ refer to fourth-order diagrams with two photons
crossed and uncrossed, respectively, and
$i=1,2,3$ refers to three consecutive lepton lines of 
the diagram of type $4a$ or $4b$.
$M_{2^*}$ is the second-order magnetic moment with a two-point vertex
insertion. $M_{2^*}[I]$ is the specific limit of $M_{2^*}$ related
to $I$-operation defined in Refs.~\cite{Cvitanovic:1974sv,Kinoshita:1990}.
The subscript $P2$ in Eq.~(\ref{aIIIyy})
means that a second-order vacuum-polarization function
$\Pi_2$ is inserted in one of photon lines
of the proper diagram in all possible ways. 
Similarly, $P2P2$ means that two $\Pi_2$'s are 
inserted in two different photon lines in all possible ways.

The numerical values of $\Delta M_{6 \alpha ,P2P2}$
are summarized in Table~\ref{tableyy}.  
Numerical values of auxiliary integrals
needed to complete the renormalization are listed in Table~\ref{tablexxaux}.

\renewcommand{\arraystretch}{0.80}
\begin{table}
\caption{ Contributions of the Set III(${\rm a}_d$) diagrams to the electron $g\!-\!2$
evaluated in {\it Version A}.
Both closed loops are electron loops.
$n_F$ is the number of Feynman diagrams represented by the integral.
All integrals are evaluated in double precision.
First 50 iterations are carried out using $1 \times 10^{8}$ sampling
points per iteration.
It is then evaluated using $1 \times 10^{9}$ sampling points 
per iteration for 50 more iterations.
\\
\label{tableyy}
}
\begin{resulttable}

$\Delta M_{6A,P2P2}$&15&-0.108~564~(~3)&$1 \times 10^8,1 \times 10^9$ &50,50 \\
$\Delta M_{6B,P2P2}$&15& 0.107~954~(11)&$1 \times 10^8,1 \times 10^9$ &50,50 \\
$\Delta M_{6C,P2P2}$&15& 0.193~333~(~6)&$1 \times 10^8,1 \times 10^9$ &50,50 \\
$\Delta M_{6D,P2P2}$&30& 0.176~456~(20)&$1 \times 10^8,1 \times 10^9$ &50,50 \\
$\Delta M_{6E,P2P2}$&15& 0.142~839~(11)&$1 \times 10^8,1 \times 10^9$ &50,50 \\
$\Delta M_{6F,P2P2}$&15& 0.194~882~(15)&$1 \times 10^8,1 \times 10^9$ &50,50 \\
$\Delta M_{6G,P2P2}$&30& 0.542~183~(34)&$1 \times 10^8,1 \times 10^9$ &50,50 \\
$\Delta M_{6H,P2P2}$&15&-0.190~026~(43)&$1 \times 10^8,1 \times 10^9$ &50,50 \\
\end{resulttable}
\end{table}
\renewcommand{\arraystretch}{1}

\renewcommand{\arraystretch}{0.80}
\begin{table}
\caption{ Auxiliary integrals for the Set III(${\rm a}_d$) and Set III(${\rm a}_s$)
with $(l_1l_2l_3) = (eee)$, where $(l_1l_2l_3)$ is defined in \ref{subsec:set3a_d}.
Six lines in the middle 
are for Set III (${\rm a}_d$) and bottom four lines are for Set III (${\rm a}_s$).
Some integrals are known exactly. 
Other integrals are obtained by VEGAS integration \cite{Lepage:1977sw}.
\\
\label{tablexxaux}
} 
\begin{renomtable}
$M_{2}$                      &  0.5 & 
$M_{2^*}$                    &  1.0  \\ 
$M_{2^*}[I]$                 & -1.0  & 
$\Delta B_{2}$               &  0.75              \\
$\Delta M_{4}$               &  0.030~833~612 ...  &
$\Delta \delta m_{4}$        &  1.906~340~(21)    \\ 
$\Delta L_{4}$               &  0.465~024~(17)    & 
$\Delta B_{4}$               & -0.437~094~(21)    \\
&   &   &  \\
$M_{2,P2}$                   &  0.015~687~421 ... & 
$M_{2^*,P2}$                 &  0.044~077~4~(3)    \\ 
$M_{2^*,P2}[I]$              &  0.010~255~3~(11)  & 
$\Delta M_{4,P2}$            & -0.106~707~082...  \\
$\Delta B_{2,P2}$            &  0.063~399~266 ... & 
$\Delta \delta m_{4,P2}$     &  0.679~769~(15)    \\ 
$\Delta L_{4,P2}$            &  0.200~092~(14)    & 
$\Delta B_{4,P2}$            & -0.314~320~(10)    \\
$\Delta M_{4,P2P2}$          & -0.026~682~(2)     & 
$\Delta \delta m_{4,P2P2}$   &  0.105~075~(11)      \\ 
$\Delta L_{4,P2P2}$          &  0.005~481~(8)      & 
$\Delta B_{4,P2P2}$          & -0.071~017~(4)      \\ 
&   &   &  \\
$M_{2,P2:2}$                 &  0.002~558~524 ... & 
$M_{2^*,P2:2}$               &  0.008~482~(1)     \\ 
$M_{2^*,P2:2}[I]$            &  0.032~904~(9)   & 
$\Delta M_{4,P2:2}$          & -0.057~587~8~(9)    \\ 
$\Delta B_{2,P2:2}$          &  0.027~902~3~(4)    &
$\Delta \delta m_{4,P2:2}$   &  0.439~326~(81)    \\ 
$\Delta L_{4,P2:2}$          &  0.094~940~(26)     & 
$\Delta B_{4,P2:2}$          & -0.199~173~(89)    \\ 
\\
\end{renomtable}
\end{table}
\renewcommand{\arraystretch}{1}

Substituting the values listed in Tables~\ref{tableyy} and \ref{tablexxaux}
into Eq.~(\ref{aIIIyy}), 
we obtain
\begin{equation}
       a_e^{(10)} [\text{Set III}({\rm a}_d): \text{\it Ver.~A}]=0.941~92~(7)  .     
\label{aIIIad_ans}
\end{equation}


\subsubsection{Electron $g\!-\!2$: Version B}
\label{subsubsec:set3a_d_electron_secondmethod}

The {\sc FORTRAN}  programs of {\it Version B} 
were generated by the automation code {\sc gencode}{\it N}
with slight modification.
Given one-line information specifying a diagram,
{\sc gencode}{\it N}  produces  a  set of programs  for a $q$-type diagram of 
any order of the perturbation theory\cite{Aoyama:2005kf,Aoyama:2007bs}.
The insertion of the vacuum polarization function 
in a  photon line is a trivial task requiring modification
of just a few lines of the {\sc gencode}{\it N} source code.
The $K$-operation method developed in  Ref.~\cite{Cvitanovic:1974sv} can
be easily automated and incorporated 
in {\sc gencode}{\it N} \cite{Aoyama:2005kf} to deal with UV divergence.
IR divergence, on the other hand, is somewhat differently treated.

The $I$-operation  defined in the previous work \cite{Kinoshita:1981vs,Kinoshita:1990} successfully  generates the IR subtraction terms  for 
a $q$-type diagram of up to the eighth-order of the perturbation theory.  
Actually, the $I$-operation works even for the tenth-order case, 
except that the automation becomes tremendously complicated. 
This is why we sought another way to handle
the IR divergence. Namely, we deviated from  the strict IR power counting, 
on which the $I$-operation is defined, and took a more diagrammatic approach.

The new scheme to deal with the IR divergence, called
$I/R$-subtraction, consists of two parts:
One is the $R$-subtraction that  removes the UV-finite part
of mass-renormalization term,
which is the cause of linear IR divergence. (The UV-divergent part
of the mass renormalization is removed by the $K$-operation.) 
Once  the mass renormalization is completed,
the remaining IR divergence is only logarithmic and is
easily subtracted  by the second part called $I$-subtraction.
This $I$-subtraction is similar to the previous $I$-operation, 
except that 
it uses  the finite part of a vertex renormalization constant 
in addition to the logarithmic IR-divergent part as an IR-counter term.
The $I/R$-subtraction can be readily incorporated 
in {\sc gencode}{\it N} \cite{Aoyama:2007bs}.

As far as the sixth-order diagrams are concerned, two methods of IR treatment, $I$-operation
or $I/R$-subtraction, work fine making no significant difference.
The difference is only finite amount in the amplitude of the magnetic moments,
which can be identified analytically.
Taking it into account, we obtain the relation of the magnetic moment
amplitudes in {\it Version A} and {\it Version B} as follows:
\begin{eqnarray}
\Delta M_{6A, P2P2}^{(B)} &=& \Delta M_{6A, P2P2}^{(A)} 
   -2\Delta L_{4b,1,P2}M_{2,P2}  -2\Delta L_{4b,1,P2P2} M_2,
\nonumber  \\
\Delta M_{6B, P2P2}^{(B)} &=& \Delta M_{6B, P2P2}^{(A)} 
   -\Delta L_{4b,2,P2}M_{2,P2}  -\Delta L_{4b,2,P2P2} M_2
\nonumber  \\
   &~~~-&\Delta \delta m_{4b,P2} (M_{2^* P2} -M_{2^* P2} [I] )  
   -\Delta \delta m_{4b,P2P2} (M_{2^* } -M_{2^*} [I] ),
\nonumber  \\
\Delta M_{6C, P2P2}^{(B)} &=& \Delta M_{6C, P2P2}^{(A)} 
   -\Delta \delta m_{4a,P2} (M_{2^* P2} -M_{2^* P2} [I] )
\nonumber  \\
   &~~~-&\Delta \delta m_{4a,P2P2} (M_{2^* } -M_{2^*} [I] ),
\nonumber  \\
\Delta M_{6D, P2P2}^{(B)} &=& \Delta M_{6D, P2P2}^{(A)} 
   -2\Delta L_{4a,1,P2}M_{2,P2}  -2\Delta L_{4a,1,P2P2} M_2,
\nonumber  \\
\Delta M_{6E, P2P2}^{(B)} &=& \Delta M_{6E, P2P2}^{(A)} 
   -\Delta L_{4a,2,P2}M_{2,P2}  -\Delta L_{4a,2,P2P2} M_2,
\nonumber  \\
\Delta M_{6F, P2P2}^{(B)} &=& \Delta M_{6F, P2P2}^{(A)}, 
\nonumber  \\
\Delta M_{6G, P2P2}^{(B)} &=& \Delta M_{6G, P2P2}^{(A)}, 
\nonumber  \\
\Delta M_{6H, P2P2}^{(B)} &=& \Delta M_{6H, P2P2}^{(A)}.
\label{substitution}
\end{eqnarray}
Note that the {\it Version B} of $\Delta M_{6\alpha, P2P2}$ 
absorbs not only $\Delta \delta m$ terms but also part of $\Delta L_4$ terms. 
From Eqs.~(\ref{aIIIyy}) and 
(\ref{substitution}) we obtain
\begin{align}
       a_e^{(10)} [\text{Set III}({\rm a}_d): \text{\it Ver.~B}]
	& = 
	\sum_{ \alpha = A}^H \Delta M_{6 \alpha ,P2P2}^{(B)}
\nonumber \\
	& -3 \Delta B_{2,P2} \Delta M_{4,P2} -3 \Delta B_2  \Delta M_{4,P2P2}
\nonumber \\
	& - [ \Delta B_{4,P2} +  \Delta L_{4,P2} - 4\Delta B_2 \Delta B_{2,P2} ] M_{2,P2}
\nonumber \\
	& - ( \Delta B_{4, P2P2} + \Delta L_{4,P2P2}-2 (\Delta B_{2,P2})^2 ) M_2 .
\label{aIIIyyy}
\end{align}
Of course this shift of terms in Eq.~(\ref{aIIIyy}) does not affect the final
result.

This is a trivial change for the Set III.
However, in Set V, which consists entirely of $q$-type tenth-order diagrams,
residual renormalization terms of $\Delta \delta m$ type
give rise to linear IR-divergences
which complicate the analysis of the renormalization scheme.
Thus there is an advantage in removing the self-mass terms
completely, not just their UV-divergent parts.

\renewcommand{\arraystretch}{0.80}
\begin{table}
\caption{ {\it Version B} contributions of the Set III(${\rm a}_d$) diagrams 
to the electron $g\!-\!2$.
The corresponding  programs are created by {\sc gencode}{\it N}.
Both closed loops are electron loops.
$n_F$ is the number of Feynman diagrams represented by the integral.
All integrals are evaluated in double precision.
\\
\label{tableyy2}
}
\begin{resulttable}
$\Delta M_{6A,P2P2}$&15&-0.130~613~(29)&$ 1 \times 10^8$ & 700 \\
$\Delta M_{6B,P2P2}$&15&-0.031~263~(26)&$ 1 \times 10^8$ & 700 \\
$\Delta M_{6C,P2P2}$&15& 0.080~981~(24)&$ 1 \times 10^8$ & 700 \\
$\Delta M_{6D,P2P2}$&30& 0.170~496~(41)&$ 1 \times 10^8$ & 700 \\
$\Delta M_{6E,P2P2}$&15& 0.183~485~(29)&$ 1 \times 10^8$ & 700 \\
$\Delta M_{6F,P2P2}$&15& 0.194~756~(31)&$ 1 \times 10^8$ & 700 \\
$\Delta M_{6G,P2P2}$&30& 0.541~900~(69)&$ 1 \times 10^8$ & 700 \\
$\Delta M_{6H,P2P2}$&15&-0.189~816~(54)&$ 1 \times 10^8$ & 700 \\
\end{resulttable}
\end{table}
\renewcommand{\arraystretch}{1}

Substituting  the values listed in Tables \ref{tablexxaux} and \ref{tableyy2} 
in Eq.~(\ref{aIIIyyy}), we obtain
\begin{equation}
       a_e^{(10)} [\text{Set III}({\rm a}_d): \text{\it Ver.~B}] = 0.941~81~(12),
\end{equation}
which is in good agreement with (\ref{aIIIad_ans}).

\subsubsection{Mass-dependent terms $A_2$ and $A_3$ of Set III(${\rm a}_d$) }
\label{subsubsec:massdependence_3ad}

Once {\sc FORTRAN} programs for mass-independent Set III(${\rm a}_d$) diagrams are obtained,
it is straightforward to evaluate contributions
of mass-dependent term $A_2^{(10)} (m_e/m_\mu)$,
$A_2^{(10)} (m_e/m_\tau)$,
and $A_3^{(10)} (m_e/m_\mu, m_e/m_\tau)$. 
We just have to choose
an appropriate fermion mass $m_{vp}$ in Eq.~(\ref{loopmass}).
Obviously the residual renormalization
terms of Set III(${\rm a}_d$) are slightly more complicated because of 
insertions of vacuum-polarization loops in several photon lines.
Note also that the integrands of these sets
may be strongly peaked because of their dependence on 
$(m_e/m_\mu)^2$ or $(m_e/m_\tau)^2$
which makes them more susceptible to the digit deficiency problem.

Of course we can evaluate them by either {\it Version A} or {\it Version B}.
Since we have established their equivalence, we may choose either one,
say {\it Version A}.

In the general case $(l_1l_2l_3)$, 
where $l_2 \neq l_3$, 
the residual renormalization terms 
of Set III(${\rm a}_d$) in {\it Version A} have the form 
\begin{align}
       a_e^{(10)} [\text{Set III}({\rm a}_d)^{(l_1l_2l_3)}] 
	& = 
	\sum_{ \alpha = A}^H \Delta M_{6 \alpha ,P2P2}^{(l_1l_2l_3)}
\nonumber \\
	& -3 \Delta B_{2,P2}^{(l_1l_2)} \Delta M_{4,P2}^{(l_1l_3)} 
	 -3 \Delta B_{2,P2}^{(l_1l_3)}\Delta M_{4,P2}^{(l_1l_2)} 
	  -3 \Delta B_2  \Delta M_{4,P2P2}^{(l_1l_2l_3)}
\nonumber \\
        & + \Delta \delta m_{4,P2}^{(l_1l_3)} ( M_{2^{*} ,P2}^{(l_1l_2)} [I] - M_{2^* , P2}^{(l_1l_2)} )
         + \Delta \delta m_{4,P2}^{(l_1l_2)} ( M_{2^{*} ,P2}^{(l_1l_3)} [I] - M_{2^* , P2}^{(l_1l_3)} )
\nonumber \\
       &  + \Delta \delta m_{4,P2P2}^{(l_1l_2l_3)} (M_{ 2^{*}} [I] -M_{2^*} ) 
\nonumber \\
	& - [ \Delta B_{4,P2}^{(l_1l_2)} + 2 \Delta L_{4,P2}^{(l_1l_2)} - 4\Delta B_2 \Delta B_{2,P2}^{(l_1l_2)} ] M_{2,P2}^{(l_1l_3)}
\nonumber \\
	& - [ \Delta B_{4,P2}^{(l_1l_3)} + 2 \Delta L_{4,P2}^{(l_1l_3)}- 4\Delta B_2 \Delta B_{2,P2}^{(l_1l_3)} ] M_{2,P2}^{(l_1l_2)}
\nonumber \\
	& - [ \Delta B_{4, P2P2}^{(l_1l_2l_3)} + 2 \Delta L_{4,P2P2}^{(l_1l_2l_3)}-4 (\Delta B_{2,P2}^{(l_1l_3)} \Delta B_{2,P2}^{(l_1l_2)} )] M_2 .
\label{aIIIyymuon}
\end{align}

For instance, for  $(l_1l_2l_3) = (eem)$,
the first, second, and third symbols 
refer to the open electron line, electron loop, and muon loop, respectively.
Some superscripts are denoted as $(l_1l_2)$ or $(l_1l_3)$
 since they have only one internal loop.
Superscripts $(l_1)$ on $\Delta B_2, M_2$, etc., are omitted for simplicity 
since these terms are mass-independent.
Note also that the second and third loops appear interchangeably 
in the case of Set III(a). Thus Eq.~(\ref{aIIIyymuon}) represents the sum
of $(eem)$ and $(eme)$.

If $l_2$ and $l_3$
represent identical particles,  duplicate terms of Eq.~(\ref{aIIIyymuon})
must be dropped to avoid double counting.

Substituting the values listed in Tables~\ref{table6} and 
\ref{tablexxauxmuon}
into Eq.~(\ref{aIIIyymuon}), 
we obtain
\begin{equation}
       a_e^{(10)} [\text{Set III}({\rm a}_d)^{(eem)}] =  0.003~166~(4) .     
\label{aIIIadmuon_ans}
\end{equation}

The contributions of diagrams of $(eet)$, $(emm)$, etc., can be calculate
by just changing the mass parameters in {\sc FORTRAN} programs.
The residual renormalization can be carried out using Eq.~(\ref{aIIIyymuon})
paying attention to whether $l_2 =l_3$ or not.
We present the final results without giving details:
\begin{eqnarray}
       a_e^{(10)} [\text{Set III}({\rm a}_d)^{(eet)}] &=& 0.000~031~55~(8) ,     
\label{aIIIadeet}  \\
       a_e^{(10)} [\text{Set III}({\rm a}_d)^{(emm)}] &=& 0.000~080~45~(11) ,    
\label{aIIIademm}  \\
       a_e^{(10)} [\text{Set III}({\rm a}_d)^{(emt)}] &=& 0.000~005~56~(2) .     
\label{aIIIademt}
\end{eqnarray}
These results have been confirmed by comparison with the results of
{\it Version B}.

The trend of mass dependence of these results
 indicates clearly that $(ett)$ case
will be an order of magnitude smaller than (\ref{aIIIademt}).
Thus it may be ignored at present.

\renewcommand{\arraystretch}{0.80}
\begin{table}
\caption{ Mass-dependent contributions of the Set III(${\rm a}_d$) diagrams to the electron $g\!-\!2$
evaluated in {\it Version A}.
One closed loop is electron loop and the other is muon loop.
Each integral is the sum of $(eem)$ and $(eme)$ types.
$n_F$ is the number of Feynman diagrams represented by the integral.
All integrals are evaluated in double precision.
\\
\label{table6}
}
\begin{resulttable}
$\Delta M_{6A,P2P2}^{(eem)}$&30&-0.000~324~(~1)&$1 \times 10^7$ &20 \\
$\Delta M_{6B,P2P2}^{(eem)}$&30& 0.000~357~(~2)&$1 \times 10^7$ &20 \\
$\Delta M_{6C,P2P2}^{(eem)}$&30& 0.000~496~(~1)&$1 \times 10^7$ &20 \\
$\Delta M_{6D,P2P2}^{(eem)}$&60& 0.000~533~(~3)&$1 \times 10^7$ &20 \\
$\Delta M_{6E,P2P2}^{(eem)}$&30& 0.000~316~(~2)&$1 \times 10^7$ &20 \\
$\Delta M_{6F,P2P2}^{(eem)}$&30& 0.000~659~(~3)&$1 \times 10^7$ &40 \\
$\Delta M_{6G,P2P2}^{(eem)}$&60& 0.001~657~(12)&$1 \times 10^7$ &60 \\
$\Delta M_{6H,P2P2}^{(eem)}$&30&-0.000~313~(12)&$1 \times 10^7$ &20 \\ 
\end{resulttable}
\end{table}
\renewcommand{\arraystretch}{1}

\subsubsection{Muon $g\!-\!2$.  Set III(${\rm a}_d$) }
\label{subsubsec:muong-2_3ad}

The leading contribution to the muon $g\!-\!2$ comes from the $(mee)$ case where
both loops consist of electrons, 
and $m$ stands for the open muon line.
Results of numerical evaluation in {\it Version A}
are listed in Table~\ref{table_mu3admee}.
From this Table and Table~\ref{tableaux_3admee} we obtain
\begin{equation}
       a_\mu^{(10)} [\text{Set III}({\rm a}_d)^{(mee)}] = 42.460~4~(188) .     
\label{aIIIad_mee}
\end{equation}
Next largest contribution comes from $(mme)$.
We list only the result:
\begin{equation}
       a_\mu^{(10)} [\text{Set III}({\rm a}_d)^{(mme)}] = 11.416~9~(22) .     
\label{aIIIad_mme}
\end{equation}
We also obtained
\begin{eqnarray}
       a_\mu^{(10)} [\text{Set III}({\rm a}_d)^{(met)}] &=& 0.421~97~(19) ,   
\label{aIIIad_met}   \\
       a_\mu^{(10)} [\text{Set III}({\rm a}_d)^{(mmt)}] &=& 0.110~71~(1),  
\label{aIIIad_mmt}  \\
       a_\mu^{(10)} [\text{Set III}({\rm a}_d)^{(mtt)}] &=& 0.008~50~(10).
\label{aIIIad_mtt}
\end{eqnarray}
These results are in good agreement with the results of {\it Version B}.

\renewcommand{\arraystretch}{0.80}
\begin{table}
\caption{ Contributions of the Set III(${\rm a}_d$) diagrams to the muon $g\!-\!2$
evaluated in {\it Version A}.
Both closed loops are electron loops.
$n_F$ is the number of Feynman diagrams represented by the integral.
All integrals are evaluated using real(10) arithmetic built in {\it gfortran}.
\\
\label{table_mu3admee}
}
\begin{resulttable}
$\Delta M_{6A,P2P2}^{(mee)}$&15&-35.558~8~(48)&$1 \times 10^7, 4 \times 10^7$ &20, 180 \\
$\Delta M_{6B,P2P2}^{(mee)}$&15& 44.427~6~(68)&$1 \times 10^7, 4 \times 10^7$ &20, 180 \\
$\Delta M_{6C,P2P2}^{(mee)}$&15& 19.208~7~(65)&$1 \times 10^7, 4 \times 10^7$ &20, 180 \\
$\Delta M_{6D,P2P2}^{(mee)}$&30& 28.117~1~(70)&$1 \times 10^7, 4 \times 10^7$ &20, 280 \\
$\Delta M_{6E,P2P2}^{(mee)}$&15& 30.980~7~(64)&$1 \times 10^7, 4 \times 10^7$ &20, ~80 \\
$\Delta M_{6F,P2P2}^{(mee)}$&15& 18.790~9~(60)&$1 \times 10^7, 4 \times 10^7$ &20, 180 \\
$\Delta M_{6g,P2P2}^{(mee)}$&30& 58.890~0~(71)&$1 \times 10^7, 4 \times 10^7$ &20, 180 \\
$\Delta M_{6h,P2P2}^{(mee)}$&15&-51.550~0~(70)&$1 \times 10^7, 4 \times 10^7$ &20, 280 \\
\end{resulttable}
\end{table}
\renewcommand{\arraystretch}{1}

\renewcommand{\arraystretch}{0.80}
\begin{table}
\caption{ Auxiliary integrals for Set III(${\rm a}_d$) and Set III(${\rm a}_s$),
where $(l_1l_2l_3) = (mee)$.
Some integrals are known exactly. 
Other integrals are obtained by VEGAS integration \cite{Lepage:1977sw}.
\\
\label{tableaux_3admee}
} 
\begin{renomtable}
$M_{2,P2}^{(me)}$                   &  1.094~275~(44)  & 
$M_{2^*,P2}^{(me)}$                 &  2.349~75~(29)    \\ 
$M_{2^*,P2}^{(me)}[I]$              & -2.183~21~(16)  & 
$\Delta B_{2,P2}^{(me)}$            &  1.885~733~(16)  \\ 
$\Delta \delta m_{4,P2}^{(me)}$     & 11.151~07~(49)    & 
$\Delta M_{4,P2}^{(me)}$            & -0.628~831~80~(2)  \\
$\Delta L_{4,P2}^{(me)}$            &  3.119~86~(66)    & 
$\Delta B_{4,P2}^{(me)}$            & -3.427~88~(49)    \\
$\Delta M_{4,P2P2}^{(mee)}$         & -1.959~37~(30)     &
$\Delta \delta m_{4,P2P2}^{(mee)}$  & 16.575~79~(52)      \\ 
$\Delta L_{4,P2P2}^{(mee)}$         &  4.960~40~(63)      & 
$\Delta B_{4,P2P2}^{(mee)}$         & -6.353~75~(62)      \\ 
&   &   &   \\
$M_{2,P2:2}^{(mee)}$                 &    2.718~651~(90)  & 
$M_{2^*,P2:2}^{(mee)}$               &    6.162~33~(39)     \\ 
$M_{2^*,P2:2}^{(mee)}[I]$            &   -5.107~35~(28)   & 
$\Delta B_{2,P2:2}^{(mee)}$          &    5.330~35~(12)    \\
$\Delta M_{4,P2:2}^{(mee)}$          &   -3.484~52~(83)    & 
$\Delta \delta m_{4,P2:2}^{(mee)}$   &   35.742~2~(12)    \\ 
$\Delta L_{4,P2:2}^{(mee)}$          &   10.621~5~(13)     & 
$\Delta B_{4,P2:2}^{(mee)}$          &  -12.811~9~(12)    \\ 
\end{renomtable}
\end{table}
\renewcommand{\arraystretch}{1}


\subsection{Set III(${\rm a}_s$)}
\label{subsec:set3a_s}

\subsubsection{Electron $g\!-\!2$.  Version A}
\label{subsubsec:set3a_s_electron_v1}

Let $M_{6 \alpha ,P2:2}$ be the magnetic moment 
projection 
of the Set III(${\rm a}_s$) generated from  self-energy-like
diagrams $6\alpha $ ($\alpha$ =A through H) 
by insertion of two $\Pi_2$'s in the same photon line
(see Fig.~\ref{fig:set3a}).
The renormalized
contribution due to these diagrams can 
be written in a way similar to Eq.~(\ref{aIIIad}). 

When summed over all the diagrams of Set III(${\rm a}_s$), the UV- and IR-divergent pieces cancel
out and the total contribution to $a^{(10)}$ can be written 
in {\it Version A} as a sum of finite pieces
(which is similar to Eq.~(5.39) of Ref.~\cite{Kinoshita:1990}):
\begin{align}
       a_e^{(10)} [\text{Set III}({\rm a}_s): \text{\it Ver.~A}]
	&= 
	\sum_{ \alpha = A}^H \Delta M_{6 \alpha ,P2:2}        
\nonumber  \\
	& -3 \Delta B_{2,P2:2} \Delta M_{4} -3 \Delta B_2  \Delta M_{4,P2:2}
\nonumber  \\
	& + \Delta \delta m_4 ( M_{2^{*} ,P2:2} [I] - M_{2^* , P2:2} ) 
	+ \Delta \delta m_{4,P2:2} (M_{ 2^{*}} [I] -M_{2^*} ) 
\nonumber  \\
	& - [ \Delta B_4 + 2 \Delta L_4 - 2( \Delta B_2 )^2 ] M_{2,P2:2}
\nonumber  \\
	& - [ \Delta B_{4, P2:2} + 2 \Delta L_{4,P2:2}-4 \Delta B_2 \Delta B_{2,P2:2} ] M_2.
\label{aIIIas}
\end{align}
The numerical values of $\Delta M_{6 \alpha ,P2:2}$        
are summarized in Table~\ref{tablexxx}.  
Numerical values of auxiliary integrals
needed to complete the renormalization are listed in Table~\ref{tablexxaux}.

\renewcommand{\arraystretch}{0.80}
\begin{table}
\caption{ {\it Version A} contributions of Set III(${\rm a}_s$) diagrams to the electron $g\!-\!2$. 
$n_F$ is the number of Feynman diagrams represented by the integral.
All integrals are evaluated in double precision.
\\
\label{tablexxx}
}
\begin{resulttable}
$\Delta M_{6A,P2:2}$&15&-0.204~682~(54)&$1 \times 10^8,1 \times 10^9$ &50,50 \\
$\Delta M_{6B,P2:2}$&15& 0.413~110~(56)&$1 \times 10^8,1 \times 10^9$ &50,50 \\
$\Delta M_{6C,P2:2}$&15& 0.458~938~(53)&$1 \times 10^8,1 \times 10^9$ &50,50 \\
$\Delta M_{6D,P2:2}$&30& 0.281~276~(54)&$1 \times 10^8,1 \times 10^9$ &50,50 \\
$\Delta M_{6E,P2:2}$&15& 0.220~637~(23)&$1 \times 10^8,1 \times 10^9$ &50,50 \\
$\Delta M_{6F,P2:2}$&15& 0.317~657~(48)&$1 \times 10^8,1 \times 10^9$ &50,50 \\
$\Delta M_{6G,P2:2}$&30& 0.765~073~(87)&$1 \times 10^8,1 \times 10^9$ &50,50 \\
$\Delta M_{6H,P2:2}$&15&-0.409~439~(98)&$1 \times 10^8,1 \times 10^9$ &50,50 \\
\end{resulttable}
\end{table}
\renewcommand{\arraystretch}{1}

Substituting the values listed in 
Tables~\ref{tablexxaux} and \ref{tablexxx}
into Eq.~(\ref{aIIIas}), we obtain
\begin{equation}
  a_e^{(10)} [\text{Set III}({\rm a}_s): \text{\it Ver.~A}] = 1.185~56~(20).     
\label{aIIIas_ans}
\end{equation}

\subsubsection{Electron $g\!-\!2$.  Version B}
\label{subsubsec:set3a_s_electron}

For the reason discussed in Sec.~\ref{subsubsec:set3a_d_electron_secondmethod}
we obtain in {\it Version B} 
a formula for $a_e^{(10)}[ \text{Set III}({\rm a}_s)]$
which is different from (\ref{aIIIas}):
\begin{align}
       a_e^{(10)} [\text{Set III}({\rm a}_s): \text{\it Ver.~B}] 
	& = 
	\sum_{ \alpha = A}^H \Delta M_{6 \alpha ,P2:2}^{(B)}
\nonumber \\
	& -3 \Delta B_{2,P2:2} \Delta M_{4} -3 \Delta B_2  \Delta M_{4,P2:2}
\nonumber \\
	& - [ \Delta B_{4} +  \Delta L_{4} - 2 (\Delta B_2)^2 ] M_{2,P2:2}
\nonumber \\
	& - ( \Delta B_{4, P2:2} + \Delta L_{4,P2:2}-4\Delta B_{2,P2:2} \Delta B_2 ) M_2,
\label{aIII(a_s)}
\end{align}
where
\begin{eqnarray}
\Delta M_{6A, P2:2}^{(B)} &=& \Delta M_{6A, P2:2}^{(A)} 
   -2\Delta L_{4b,1}M_{2,P2:2}  -2\Delta L_{4b,1,P2:2} M_2,
\nonumber  \\
\Delta M_{6B, P2:2}^{(B)} &=& \Delta M_{6B, P2:2}^{(A)} 
   -\Delta L_{4b,2}M_{2,P2:2}  -\Delta L_{4b,2,P2:2} M_2
\nonumber  \\
   &~~~-&\Delta \delta m_{4b} (M_{2^* P2:2} -M_{2^* P2:2} [I] )  
   -\Delta \delta m_{4b,P2:2} (M_{2^* } -M_{2^*} [I] ),
\nonumber  \\
\Delta M_{6C, P2:2}^{(B)} &=& \Delta M_{6C, P2:2}^{(A)} 
   -\Delta \delta m_{4a} (M_{2^* P2:2} -M_{2^* P2:2} [I] )
\nonumber  \\
   &~~~-&\Delta \delta m_{4a,P2:2} (M_{2^* } -M_{2^*} [I] ),
\nonumber  \\
\Delta M_{6D, P2:2}^{(B)} &=& \Delta M_{6D, P2:2}^{(A)} 
   -2\Delta L_{4a,1}M_{2,P2:2}  -2\Delta L_{4a,1,P2:2} M_2,
\nonumber  \\
\Delta M_{6E, P2:2}^{(B)} &=& \Delta M_{6E, P2:2}^{(A)} 
   -\Delta L_{4a,2}M_{2,P2:2}  -\Delta L_{4a,2,P2:2} M_2,
\nonumber  \\
\Delta M_{6F, P2:2}^{(B)} &=& \Delta M_{6F, P2:2}^{(A)}, 
\nonumber  \\
\Delta M_{6G, P2:2}^{(B)} &=& \Delta M_{6G, P2:2}^{(A)}, 
\nonumber  \\
\Delta M_{6H, P2:2}^{(B)} &=& \Delta M_{6H, P2:2}^{(A)}.
\label{substitution_as}
\end{eqnarray}

From Tables \ref{tablexxaux} and \ref{tableyy3} 
we obtain

\begin{equation}
       a_e^{(10)} [\text{Set III}({\rm a}_s): \text{\it Ver.~B}] = 1.185~26~(24),
\end{equation}
in good agreement with (\ref{aIIIas_ans}).

\renewcommand{\arraystretch}{0.80}
\begin{table}
\caption{ {\it Version B} contributions of 
Set III(${\rm a}_s$) diagrams to the electron $g\!-\!2$.
Programs are created by {\sc gencode}{\it N}.
$n_F$ is the number of Feynman diagrams represented by the integral.
All integrals are evaluated in double precision.
\\
\label{tableyy3}
}
\begin{resulttable}
$\Delta M_{6A,P2:2}$&15&-0.280~605~(31)& $ 1 \times 10^8$ & 600 \\
$\Delta M_{6B,P2:2}$&15&-0.068~830~(70)& $ 1 \times 10^8$ & 600 \\
$\Delta M_{6C,P2:2}$&15& 0.070~002~(67)& $ 1 \times 10^8$ & 600 \\
$\Delta M_{6D,P2:2}$&30& 0.269~937~(106)&$ 1 \times 10^8$ & 600 \\
$\Delta M_{6E,P2:2}$&15& 0.297~943~(60)& $ 1 \times 10^8$ & 600 \\
$\Delta M_{6F,P2:2}$&15& 0.317~432~(69)& $ 1 \times 10^8$ & 600 \\
$\Delta M_{6G,P2:2}$&30& 0.764~711~(127)&$ 1 \times 10^8$ & 600 \\
$\Delta M_{6H,P2:2}$&15&-0.409~110~(103)&$ 1 \times 10^8$ & 600 \\
\end{resulttable}
\end{table}
\renewcommand{\arraystretch}{1}

\subsubsection{Mass-dependent terms $A_2$ and $A_3$ of Set III(${\rm a}_s$) }
\label{subsubsec:massdependence_3as}

For the Set III(${\rm a}_s$) we have (in {\it Version A})
\begin{align}
       a_e^{(10)} [\text{Set III}({\rm a}_s)^{(l_1 l_2 l_3)}] 
	&= 
	\sum_{ \alpha = A}^H \Delta M_{6 \alpha ,P2:2}^{(l_1 l_2 l_3)}        
\nonumber  \\
	& -3 \Delta B_{2,P2:2}^{(l_1 l_2 l_3)} \Delta M_{4} -3 \Delta B_2  \Delta M_{4,P2:2}^{(l_1 l_2 l_3)}
\nonumber  \\
	& + \Delta \delta m_4 ( M_{2^{*} ,P2:2} [I]^{(l_1 l_2 l_3)} - M_{2^* , P2:2}^{(l_1 l_2 l_3)} ) 
	+ \Delta \delta m_{4,P2:2}^{(l_1 l_2 l_3)} (M_{ 2^{*}} [I] -M_{2^*} ) 
\nonumber  \\
	& - [ \Delta B_4 + 2 \Delta L_4 - 2( \Delta B_2 )^2 ] M_{2,P2:2}^{(l_1 l_2 l_3)}
\nonumber  \\
	& - [ \Delta B_{4, P2:2}^{(l_1 l_2 l_3)}+ 2 \Delta L_{4,P2:2}^{(l_1 l_2 l_3)}-4 \Delta B_2 \Delta B_{2,P2:2}^{(l_1 l_2 l_3)} ] M_2.
\label{aIIIasmuon}
\end{align}

\renewcommand{\arraystretch}{0.80}
\begin{table}
\caption{ Contributions of diagrams of Set III(${\rm a}_s$)
containing one 
electron vacuum-polarization loop and one
muon vacuum-polarization loop.
$n_F$ is the number of Feynman diagrams represented by the integral.
All integrals are evaluated in double precision.
\\
\label{table7}
}
\begin{resulttable}
$\Delta M_{6A,P2:2}^{(eem)}$&30&-0.000~541~(~4)&$1 \times 10^7$ &20 \\
$\Delta M_{6B,P2:2}^{(eem)}$&30& 0.001~434~(17)&$1 \times 10^7$ &20 \\
$\Delta M_{6C,P2:2}^{(eem)}$&30& 0.001~334~(26)&$1 \times 10^7$ &20 \\
$\Delta M_{6D,P2:2}^{(eem)}$&60& 0.000~747~(26)&$1 \times 10^7$ &20 \\
$\Delta M_{6E,P2:2}^{(eem)}$&30& 0.000~494~(10)&$1 \times 10^7$ &20 \\
$\Delta M_{6F,P2:2}^{(eem)}$&30& 0.001~233~(16)&$1 \times 10^7$ &20 \\
$\Delta M_{6G,P2:2}^{(eem)}$&60& 0.002~397~(78)&$1 \times 10^7$ &20 \\
$\Delta M_{6H,P2:2}^{(eem)}$&30&-0.000~995~(43)&$1 \times 10^7$ &20 \\ 
\end{resulttable}
\end{table}
\renewcommand{\arraystretch}{1}

\renewcommand{\arraystretch}{0.80}
\begin{table}
\caption{ Contributions of the Set III(${\rm a}_s$) diagrams to the muon $g\!-\!2$
evaluated in {\it Version A}.
Both closed loops are electron loops.
$n_F$ is the number of Feynman diagrams represented by the integral.
All integrals are evaluated using real(10) arithmetic built in {\it gfortran}.
\\
\label{table_mu3asmee}
}
\begin{resulttable}
$\Delta M_{6A,P2P2}^{(mee)}$&15&-38.157~8~(106)&$1 \times 10^7, 4 \times 10^7$ &20,  80 \\
$\Delta M_{6B,P2P2}^{(mee)}$&15& 51.107~5~(103)&$1 \times 10^7, 4 \times 10^7$ &20, 180 \\
$\Delta M_{6C,P2P2}^{(mee)}$&15& 20.547~7~(113)&$1 \times 10^7, 4 \times 10^7$ &20, 180 \\
$\Delta M_{6D,P2P2}^{(mee)}$&30& 31.469~8~(113)&$1 \times 10^7, 4 \times 10^7$ &20, 280 \\
$\Delta M_{6E,P2P2}^{(mee)}$&15& 32.799~0~(89)&$1 \times 10^7, 4 \times 10^7$ &20,  80 \\
$\Delta M_{6F,P2P2}^{(mee)}$&15& 19.014~4~(96)&$1 \times 10^7, 4 \times 10^7$ &20, 180 \\
$\Delta M_{6G,P2P2}^{(mee)}$&30& 61.519~4~(101)&$1 \times 10^7, 4 \times 10^7$ &20, 180 \\
$\Delta M_{6H,P2P2}^{(mee)}$&15&-55.142~4~( 95)&$1 \times 10^7, 4 \times 10^7$ &20, 280 \\
\end{resulttable}
\end{table}
\renewcommand{\arraystretch}{1}

\renewcommand{\arraystretch}{0.80}
\begin{table}
\caption{ Auxiliary integrals
which depend on the mass ratio $m_\mu/m_e$.
Those for Set III(${\rm a}_d$) are listed on the left side.
Those for Set III(${\rm a}_s$) are listed on the top half of the right side.
Those for Set III(b) are listed on the bottom half of the right side.
\\
\label{tablexxauxmuon}
} 
\begin{renomtable}
$\Delta M_{4,P2}^{(em)}$     & -0.000~018~9~(~1)  &
$\Delta M_{4,P2:2}^{(eem)}$  & -0.000~074~7~(1)  \\
$\Delta M_{4,P2P2}^{(eem)}$  & -0.000~019~1~(~~1)  &
$M_{2^*,P2:2}^{(eem)}$       &  0.000~000~9~(0)     \\ 
$M_{2,P2}^{(em)}$            &  0.015~690~0~(~16) & 
$M_{2^*,P2:2}[I]^{(eem)}$    &  0.000~069~7~(1)     \\ 
$M_{2^*,P2}^{(em)}$          &  0.044~089~4~(101)    &
$\Delta B_{2,P2:2}^{(eem)}$  &  0.000~036~2~(1)    \\
$M_{2^*,P2}[I]^{(em)}$       &  0.010~274~2~(256)   & 
$\Delta B_{4,P2:2}^{(eem)}$  & -0.000~409~7~(~48)   \\
$\Delta B_{2,P2}^{(em)}$     &  0.000~009~4~(00)  & 
$\Delta L_{4,P2:2}^{(eem)}$  &  0.000~016~0~(~59)   \\
$\Delta B_{4,P2}^{(em)}$     & -0.000~091~5~(~4)   &
$\Delta \delta m_{4,P2:2}^{(eem)}$   &  0.001~261~2~(63)      \\ 
$\Delta L_{4,P2}^{(em)}$     &  0.000~012~7~(~6)    & 
          &   \\
$\Delta B_{4,P2P2}^{(eem)}$  & -0.000~129~0~(~3)      & 
$\Delta M_{4,P4}^{(em)}$       & -0.000~068~2~(~7)   \\
$\Delta L_{4,P2P2}^{(eem)}$  & -0.000~113~6~(~6)      & 
$M_{2^*,P4}^{(em)}$          &  0.000~005~9~(0)    \\
$\Delta \delta m_{4,P2}^{(em)}$  &  0.000~253~9~(~5)    & 
$M_{2^*,P4}[I]^{(em)}$       &  0.000~052~0~(~1)    \\
$\Delta \delta m_{4,P2P2}^{(eem)}$   &  0.000~195~1~(~3)    & 
$\Delta B_{4,P4}^{(em)}$     & -0.000~322~0~(~9)   \\
     &     &
$\Delta L_{4,P4}^{(em)}$     &  0.000~054~2~(13)   \\
      &     &
$\Delta \delta m_{4,P4}^{(em)}$  &  0.000~878~6~(15)   \\
       &    &    &   \\
\end{renomtable}
\end{table}
\renewcommand{\arraystretch}{1}

Substituting the values listed in Tables~\ref{table7} and 
\ref{tablexxauxmuon}
into Eq.~(\ref{aIIIasmuon}), 
we obtain
\begin{equation}
       a_e^{(10)} [\text{Set III}({\rm a}_s)^{(eem)}] 
	=0.004~12~(10) .     
\label{aIIIasmuon_eem}
\end{equation}
We also obtained
\begin{eqnarray}
       a_e^{(10)} [\text{Set III}({\rm a}_s)^{(emm)}] 
	&=& 0.000 ~144 ~7~ (9),    
\label{aIIIasmuon_emm}   \\
       a_e^{(10)} [\text{Set III}({\rm a}_s)^{(eet)}] 
	&=& 0.000 ~045 ~95~ (23),    
\label{aIIIasmuon_eet}   \\
       a_e^{(10)} [\text{Set III}({\rm a}_s)^{(emt)}] 
	&=& 0.000~009~88~(8).     
\label{aIIIasmuon_emt}
\end{eqnarray}
These results are in good agreement with those of {\it Version B}.
The contribution of the $(ett)$ term is negligibly small. 

\subsubsection{Muon $g\!-\!2$.  Set III(${\rm a}_s$) }
\label{subsubsec:muong-2_3as}

The leading contribution to the muon $g\!-\!2$ comes from the case where
both loops consist of electrons, namely the $(mee)$ case,
where $m$ stands for the muon.
Results of numerical evaluation in {\it Version A}
are listed in Table \ref{table_mu3asmee}.
From this Table and Table \ref{tableaux_3admee} we obtain
\begin{equation}
       a_\mu^{(10)} [\text{Set III}({\rm a}_s)^{(mee)}] = 43.048~8~(194) .     
\label{aIIIas_mee}
\end{equation}
Next leading term is
\begin{equation}
       a_\mu^{(10)} [\text{Set III}({\rm a}_s)^{(mem)}] = 12.190~3~(176) .     
\label{aIIIas_mem}
\end{equation}
We also have
\begin{eqnarray}
       a_\mu^{(10)} [\text{Set III}({\rm a}_s)^{(met)}] &=& 0.469~43~(39),   
\label{aIIIas_met}   \\
       a_\mu^{(10)} [\text{Set III}({\rm a}_s)^{(mmt)}] &=& 0.150~11~(21) ,    
\label{aIIIas_mmt}   \\
       a_\mu^{(10)} [\text{Set III}({\rm a}_s)^{(mtt)}] &=&  0.013~68~(3).     
\label{aIIIas_mtt}
\end{eqnarray}
These results are in good agreement with those of {\it Version B}.

\section{Set III(b)}
\label{sec:set3b}

\begin{figure}[tb]
\includegraphics[width=6cm]{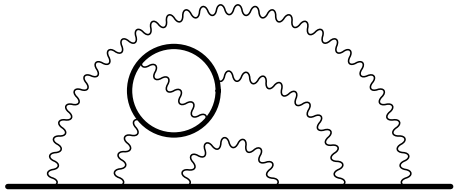}
\caption{\label{fig:set3b} 
Typical tenth-order diagrams of Set III(b) obtained by insertion of
a fourth-order vacuum-polarization loop $\Pi_4$ in lepton 
diagrams of the three-photon-exchange type.  Altogether there are
450 diagrams of this type.
}
\end{figure}

Diagrams belonging to this set are generated 
by inserting a proper fourth-order
vacuum-polarization loop  $\Pi_4$ (consisting of three diagrams) in
the photon lines of  $M_6$.
Time-reversal invariance and use of the photon spectral function 
${\rho}_4$ reduce the
number of independent integrals from 450 to 8.  
These integrals are represented by the 
``self-energy-like'' diagrams of Fig.~\ref{fig:M6}.
A typical diagram is shown in Fig.~\ref{fig:set3b}.

\setcounter{subsubsection}{0}
\subsubsection{Electron $g\!-\!2$: Version A}
\label{subsubsec:set3b_electron_firstmethod}

Let $M_{6 \alpha ,P4}$ be the magnetic moment 
projection 
of the set of diagrams generated from a self-energy
diagram $ \alpha $ (=A through H) of Fig.~\ref{fig:M6} 
by insertion of $\Pi_4$ and
an external vertex.
The renormalized
contribution due to the Set III(b) diagrams can then 
be written as 
\begin{equation}
       a_e^{(10)} [\text{Set III}({\rm b}): \text{\it Ver.~A}] =\sum_{\alpha = A}^H  a_{6 \alpha ,P4} ,
\label{aIIIb}
\end{equation}
with
\begin{equation}
	a_{6 \alpha ,P4} 
	= \Delta M_{6 \alpha ,P4} + \text{residual renormalization terms} ,
\end{equation}
where all divergences, except those within $\Pi_4$,
are removed by intermediate renormalization by $ K_S$ and
$I_R$ operations.  (See Ref.~\cite{Kinoshita:1990}.)  

The numerical values
of Set III(b) integrals are summarized in 
Table~\ref{table5}.  
Numerical values of auxiliary integrals
needed to complete the renormalization are listed in Table~\ref{table5aux}.

When summed over all the diagrams of Set III(b), the UV- and IR-divergent pieces cancel
out and the total contribution to $a^{(10)}$ can be written as a sum of
finite pieces
(which is similar to Eq.~(5.39) of Ref.~\cite{Kinoshita:1990}):
\begin{align}
       a_e^{(10)} [\text{Set III}({\rm b}): \text{\it Ver.~A}] 
	&= 
	\sum_{ \alpha = A}^H \Delta M_{6 \alpha ,P4}        
\nonumber  \\
	& -3 \Delta B_{2,P4} \Delta M_{4} -3 \Delta B_2  \Delta M_{4,P4}
\nonumber  \\
	& + \Delta \delta m_4 ( M_{2^{*} ,P4} [I] - M_{2^* , P4} ) 
	+ \Delta \delta m_{4,P4} (M_{ 2^{*}} [I] -M_{2^*} )  
\nonumber  \\
	& - [ \Delta B_4 + 2 \Delta L_4 - 2( \Delta B_2 )^2 ] M_{2,P4} 
\nonumber  \\
	& - [ \Delta B_{4, P4} + 2 \Delta L_{4,P4}-4 \Delta B_2 \Delta B_{2,P4} ] M_2 .
\label{aIIIA}
\end{align}
Terms with suffix $P_4$ in Eq.~(\ref{aIIIA}) 
are obtained by insertion of $\Pi_4$
in the photon lines of diagrams.
Note that $K$-operation is not applied to $\Pi_4$ so that
we have $M_{2,P4}$, instead of $\Delta M_{2,P4}$, in Eq.~(\ref{aIIIA}). 

\renewcommand{\arraystretch}{0.80}
\begin{table}
\caption{ {\it Version A} contributions of 
Set III(b) diagrams to the electron $g\!-\!2$.
$n_F$ is the number of Feynman diagrams represented by the integral.
All integrals are evaluated in double precision.
\\
\label{table5}
}
\begin{resulttable}
$\Delta M_{6A,P4}$&15&-1.275~23~(~8)&$1 \times 10^8,1 \times 10^9$ &50,50 \\
$\Delta M_{6B,P4}$&15& 1.865~05~(14)&$1 \times 10^8,1 \times 10^9$ &50,50 \\
$\Delta M_{6C,P4}$&15& 1.593~72~(14)&$1 \times 10^8,1 \times 10^9$ &50,50 \\
$\Delta M_{6D,P4}$&30& 1.166~99~(14)&$1 \times 10^8,1 \times 10^9$ &50,50 \\
$\Delta M_{6E,P4}$&15& 1.212~50~(~6)&$1 \times 10^8,1 \times 10^9$ &50,50 \\
$\Delta M_{6F,P4}$&15& 1.113~25~(14)&$1 \times 10^8,1 \times 10^9$ &50,50 \\
$\Delta M_{6G,P4}$&30& 2.948~70~(24)&$1 \times 10^8,1 \times 10^9$ &50,50 \\
$\Delta M_{6H,P4}$&15&-2.231~76~(22)&$1 \times 10^8,1 \times 10^9$ &50,50 \\ 
\end{resulttable}
\end{table}
\renewcommand{\arraystretch}{1}

\renewcommand{\arraystretch}{0.80}
\begin{table}
\caption{ Auxiliary integrals for the Set III(b).
Some integrals are known exactly. 
Other integrals are obtained by VEGAS integration.
\\
\label{table5aux}
} 
\begin{renomtable}
$M_{2}$                          &  0.5  &
$M_{2,P4}$                       &  0.052~870~652 ...  \\
$M_{2^*}$                        &  1.0  &
$M_{2^{*},P4}$                   &  0.145~597~(21)  \\
$M_{2^*}[I]$                     & -1.0  &
$M_{2^*,P4}[I]$                  & -0.016~526~(69)  \\
$\Delta M_4$                     &  0.030~833~612 ...  &
$\Delta M_{4,P4}$                & -0.288~997~(12)  \\
$\Delta \delta m_4$              &  1.906~340~(21)  &
$\Delta \delta m_{4,P4}$         &  1.773~79~(26)  \\
$\Delta B_2$                     &  0.75  &
$\Delta B_{2,P4}$                &  0.183~666~8~(18)  \\
$\Delta B_4$                     & -0.437~094~(21)  &
$\Delta B_{4,P4}$                & -0.816~23~(25)  \\
$\Delta L_4$                     &  0.465~024~(17)  &
$\Delta L_{4,P4}$                &  0.559~72~(25)  \\
\end{renomtable}
\end{table}
\renewcommand{\arraystretch}{1}

Substituting the values listed 
in Tables~\ref{table5} and \ref{table5aux}
into Eq.~(\ref{aIIIA}), we obtain
\begin{equation}
       a_e^{(10)} [\text{Set III}({\rm b}): \text{\it Ver.~A}]=3.327~14~(56)  .     
\label{aIIIbA}
\end{equation}


\subsubsection{Electron $g\!-\!2$: Version B}
\label{subsubsec:set3b_electron_secondmethod}

Let us now treat Set III(b) by the method
based on the automated code generation scheme.
In this approach, the contribution from Set III(b) is
expressed as
\begin{align}
       a_e^{(10)} [\text{Set III}({\rm b}): \textit{\it Ver.~B}] 
	&= 
	\sum_{\alpha=A}^{H} \Delta M_{6\alpha,P4}^{(B)}
\nonumber \\
	& -3 \Delta B_2 \Delta M_{4,P4}   
          -  3 \Delta B_{2,P4} \Delta M_4   
\nonumber \\
	& -(\Delta B_{4} +\Delta L_{4}
	        -2(\Delta B_{2})^2 ) M_{2,P4}
\nonumber \\
	& - ( \Delta B_{4,P4} +\Delta L_{4,P4}
	                   - 4\Delta B_2 \Delta B_{2,P4} ) M_2,
	\label{a_III(b)_method_B}
\end{align}
where
\begin{eqnarray}
\Delta M_{6A, P4}^{(B)} &=& \Delta M_{6A, P4}^{(A)} 
   -2\Delta L_{4b,1}M_{2,P4}  -2\Delta L_{4b,1,P4} M_2,
\nonumber  \\
\Delta M_{6B, P4}^{(B)} &=& \Delta M_{6B, P4}^{(A)} 
   -\Delta L_{4b,2}M_{2,P4}  -\Delta L_{4b,2,P4} M_2
\nonumber  \\
   &~~~-&\Delta \delta m_{4b} (M_{2^* P4} -M_{2^* P4} [I] )  
   -\Delta \delta m_{4b,P4} (M_{2^* } -M_{2^*} [I] ),
\nonumber  \\
\Delta M_{6C, P4}^{(B)} &=& \Delta M_{6C, P4}^{(A)} 
   -\Delta \delta m_{4a} (M_{2^* P4} -M_{2^* P4} [I] )
\nonumber  \\
   &~~~-&\Delta \delta m_{4a,P4} (M_{2^* } -M_{2^*} [I] ),
\nonumber  \\
\Delta M_{6D, P4}^{(B)} &=& \Delta M_{6D, P4}^{(A)} 
   -2\Delta L_{4a,1}M_{2,P4}  -2\Delta L_{4a,1,P4} M_2,
\nonumber  \\
\Delta M_{6E, P4}^{(B)} &=& \Delta M_{6E, P4}^{(A)} 
   -\Delta L_{4b,2}M_{2,P4}  -\Delta L_{4b,2,P4} M_2,
\nonumber  \\
\Delta M_{6F, P4}^{(B)} &=& \Delta M_{6F, P4}^{(A)}, 
\nonumber  \\
\Delta M_{6G, P4}^{(B)} &=& \Delta M_{6G, P4}^{(A)}, 
\nonumber  \\
\Delta M_{6H, P4}^{(B)} &=& \Delta M_{6H, P4}^{(A)}.
\label{substitution_b}
\end{eqnarray}

Using the code generator we obtained the programs of  the magnetic moments
$M_{6\alpha, P4}$, $\alpha=A,\dots, H$, 
and 
$M_{4\alpha}$, 
$M_{4{\alpha},P4}$, $\alpha=A,B$. 
The programs for the renormalization constants 
$L_{4\alpha,P4}$, 
$L_{4\alpha}$, 
$B_{4\alpha,P4}$, 
$B_{4\alpha}$, 
$\delta m_{4\alpha,P4}$, 
$\delta m_{4\alpha}$ 
are also automatically generated. 
Other quantities, 
$\Delta B_{2,P4}$, 
$M_{2,P4}$ 
are  very simple so that they are calculated by using hand-written programs. 
The values of $\Delta B_2$ and $M_2$ are analytically known. 

The results of numerical integration by VEGAS are shown in 
Table \ref{M6_without_pms}.  
\renewcommand{\arraystretch}{0.80}
\begin{table}
\caption{ {\it Version B} contributions of  Set III(b) diagrams to the electron $g\!-\!2$. 
Programs are created by {\sc gencode}{\it N}.
$n_F$ is the number of Feynman diagrams represented by the integral.
All integrals are evaluated in double precision.
\label{M6_without_pms}
}
\begin{resulttable}
$\Delta M_{6A,P4}$& 15 & -1.673~94~(34) &  $1\times 10^8,1\times 10^9$ & 50, 50 \\
$\Delta M_{6B,P4}$& 15 & -0.959~45~(24) &  $1\times 10^8,1\times 10^9$ & 50, 50 \\
$\Delta M_{6C,P4}$& 15 &  0.427~34~(21) &  $1\times 10^8,1\times 10^9$ & 50, 50 \\
$\Delta M_{6D,P4}$& 30 &  1.110~10~(32) &  $1\times 10^8,1\times 10^9$ & 50, 50 \\
$\Delta M_{6E,P4}$& 15 &  1.498~03~(18) &  $1\times 10^8,1\times 10^9$ & 50, 50 \\
$\Delta M_{6F,P4}$& 15 &  1.113~12~(20) &  $1\times 10^8,1\times 10^9$ & 50, 50 \\
$\Delta M_{6G,P4}$& 30 &  2.947~48~(36) &  $1\times 10^8,1\times 10^9$ & 50, 50 \\
$\Delta M_{6H,P4}$& 15 & -2.231~66~(28) &  $1\times 10^8,1\times 10^9$ & 50, 50 \\
\end{resulttable}
\end{table}
\renewcommand{\arraystretch}{1}


Substituting the numbers shown 
in Tables~\ref{M6_without_pms} and \ref{table5aux}
into Eq.~(\ref{a_III(b)_method_B}), we obtain
\begin{equation}
       a_e^{(10)} [\text{Set III}({\rm b}): \textit{\it Ver.~B}] = 3.327~07~(78)
\end{equation}
in good agreement with (\ref{aIIIbA}),
where the uncertainty is from the numerical integration only.

\subsubsection{Mass-dependent terms $A_2$ of Set III(b) }
\label{subsubsec:massdependence_3b}

The residual renormalization scheme 
(in {\it Version A}) for the $(em)$ term is the
following:
\begin{align}
       a_e^{(10)} [\text{Set III}({\rm b})^{(em)}] 
	&= 
	\sum_{ \alpha = A}^H \Delta M_{6 \alpha ,P4}^{(em)}        
\nonumber  \\
	& -3 \Delta B_{2,P4}^{(em)} \Delta M_{4} -3 \Delta B_2  \Delta M_{4,P4}^{(em)}
\nonumber  \\
	& + \Delta \delta m_4 ( M_{2^{*} ,P4} [I]^{(em)} - M_{2^* , P4}^{(em)} ) 
	+ \Delta \delta m_{4,P4}^{(em)}(M_{ 2^{*}} [I] -M_{2^*} )  
\nonumber  \\
	& - [ \Delta B_4 + 2 \Delta L_4 - 2( \Delta B_2 )^2 ] M_{2,P4}^{(em)} 
\nonumber  \\
	& - [ \Delta B_{4, P4}^{(em)} + 2 \Delta L_{4,P4}^{(em)}-4 \Delta B_2 \Delta B_{2,P4}^{(em)} ] M_2 .
\label{aIIIAmuon}
\end{align}

Substituting the values listed in 
Tables~\ref{tablexxauxmuon} and \ref{table8}  
into Eq.~(\ref{aIIIAmuon}), 
we obtain
\begin{equation}
       a_e^{(10)} [\text{Set III}({\rm b})^{(em)}] 
	=0.002~794~(1) .     
\label{aIIIb_ans_em}
\end{equation}

It is easy to obtain the contribution of tau lepton loop
instead of the muon loop.
We have simply to replace the muon mass by the tau mass
in the {\sc FORTRAN} programs.
For instance a crude calculation in {\it Version A} yields
\begin{equation}
       a_e^{(10)} [\text{Set III}({\rm b})^{(et)}] 
	=0.000~021~42~(1) ,     
\label{aIIIb_ans_et}
\end{equation}
%
which is two orders of magnitude smaller than (\ref{aIIIb_ans_em}).

\subsubsection{Muon $g\!-\!2$.  Set III(b) }
\label{subsubsec:muong-2_3b}

The leading contribution to the muon $g\!-\!2$ comes from the case 
containing an electron loop, namely the $(me)$ case,
where $m$ stands for the muon.
Results of numerical evaluation ({\it Version A})
are listed in Table \ref{table_mu3bme}.
From this Table and Table \ref{table3s_me_aux} we obtain
\begin{equation}
       a_\mu^{(10)} [\text{Set III}({\rm b})^{(me)}] = 11.936~7~(45).     
\label{aIIIb_me}
\end{equation}
We also obtained ({\it Version A})
\begin{equation}
       a_\mu^{(10)} [\text{Set III}({\rm b})^{(mt)}] = 0.143~60~(1).     
\label{aIIIb_mt}
\end{equation}
These results are confirmed by {\it Version B} calculation.

\renewcommand{\arraystretch}{0.80}
\begin{table}
\caption{ Contributions of $(em)$  diagrams of Fig.~\ref{fig:set3}
to the Set III(b).
$n_F$ is the number of Feynman diagrams represented by the integral.
All integrals are evaluated in double precision.
\\
\label{table8}
}
\begin{resulttable}
$\Delta M_{6A,P4}^{(em)}$&15&-0.000~389~(~2)&$1 \times 10^7$ &20 \\
$\Delta M_{6B,P4}^{(em)}$&15& 0.000~943~(~6)&$1 \times 10^7$ &20 \\
$\Delta M_{6C,P4}^{(em)}$&15& 0.000~925~(~6)&$1 \times 10^7$ &20 \\
$\Delta M_{6D,P4}^{(em)}$&30& 0.000~592~(~4)&$1 \times 10^7$ &20 \\
$\Delta M_{6E,P4}^{(em)}$&15& 0.000~371~(~4)&$1 \times 10^7$ &20 \\
$\Delta M_{6F,P4}^{(em)}$&15& 0.000~773~(~7)&$1 \times 10^7$ &40 \\
$\Delta M_{6G,P4}^{(em)}$&30& 0.001~626~(26)&$1 \times 10^7$ &60 \\
$\Delta M_{6H,P4}^{(em)}$&15&-0.000~751~(20)&$1 \times 10^7$ &20 \\ 
\end{resulttable}
\end{table}
\renewcommand{\arraystretch}{1}

\renewcommand{\arraystretch}{0.80}
\begin{table}
\caption{Contributions of $(me)$-type diagrams of Set III(b) to the muon $g\!-\!2$.
$n_F$ is the number of Feynman diagrams represented by the integral.
All integrals are evaluated using the real(10) arithmetic 
built in {\it gfortran}.
\\
\label{table_mu3bme}
}
\begin{resulttable}
$\Delta M_{6A,P4}^{(me)}$&15& -16.0005~(53)&$1 \times 10^7$ &  600 \\
$\Delta M_{6B,P4}^{(me)}$&15&  25.3670~(61)&$1 \times 10^7$ &  900 \\
$\Delta M_{6C,P4}^{(me)}$&15&   2.8871~(61)&$1 \times 10^7$ &  900 \\
$\Delta M_{6D,P4}^{(me)}$&30&  12.8166~(79)&$1 \times 10^7$ & 1000 \\
$\Delta M_{6E,P4}^{(me)}$&15&  13.6292~(44)&$1 \times 10^7$ &  600 \\
$\Delta M_{6F,P4}^{(me)}$&15&   6.9195~(59)&$1 \times 10^7$ &  600 \\
$\Delta M_{6G,P4}^{(me)}$&30&  24.6728~(59)&$1 \times 10^7$ &  700 \\
$\Delta M_{6H,P4}^{(me)}$&15& -23.0066~(57)&$1 \times 10^7$ &  800 \\
\end{resulttable}
\end{table}
\renewcommand{\arraystretch}{1}

\renewcommand{\arraystretch}{0.80}
\begin{table}
\caption{ Auxiliary integrals for the Set III$({\rm b})^{(me)}$.
Some integrals are known exactly. 
Other integrals are obtained by VEGAS integration.
\\
\label{table3s_me_aux}
} 
\begin{renomtable}
$M_{2}$                          &  0.5  &
$M_{2,P4}^{(me)}$                       &  1.493~651~(84)   \\
$M_{2^*}$                        &  1.0  &
$M_{2^{*},P4}^{(me)}$                   &  3.122~88~(16)  \\
$M_{2^*}[I]$                     & -1.0  &
$M_{2^*,P4}^{(me)}[I]$                  & -2.996~76~(24)  \\
$\Delta M_4$                     &  0.030~833~612 ...  &
$\Delta M_{4,P4}^{(me)}$                & -0.438~76~(26)  \\
$\Delta \delta m_4$              &  1.906~340~(21)  &
$\Delta \delta m_{4,P4}^{(me)}$         & 13.651~22~(80)  \\
$\Delta B_2$                     &  0.75  &
$\Delta B_{2,P4}^{(me)}$                &  2.439~109~(53)  \\
$\Delta B_4$                     & -0.437~094~(21)  &
$\Delta B_{4,P4}^{(me)}$                & -3.826~32~(71)  \\
$\Delta L_4$                     &  0.465~024~(17)  &
$\Delta L_{4,P4}^{(me)}$                &  3.653~31~(42)  \\
\end{renomtable}
\end{table}
\renewcommand{\arraystretch}{1}

\section{Discussion}
\label{sec:discussion}

As was noted earlier {\it Version A} and {\it Version B}
differ in the treatment
of self-energy subtraction and IR divergence.
Furthermore, the actual algebraic form of integrands in the first
method \cite{Cvitanovic:1974uf} is quite different from the second one
because "Kirchhoff's laws" satisfied by the scalar currents \cite{Kinoshita:1990}
were used extensively to make the integrand as compact as possible
to save the computing time. 
Thus the two calculations can be regarded as independent of each
other and the results agree within their error bars.
Thus they may be combined statistically
to yield the values listed below:
\begin{eqnarray}
       a_e^{(10)} [\text{Set III}({\rm a}_d)] 
 &=& 0.9419~(1),
\label{3adeee}
 \\
       a_e^{(10)} [\text{Set III}({\rm a}_s)] 
 &=& 1.1854~(2),
\label{3aseee}
 \\
       a_e^{(10)} [\text{Set III}({\rm b})] 
 &=& 3.3271~(5).
\label{3beee}
\end{eqnarray}

The mass-dependent contribution of Set III(${\rm a}_d$)
to the electron $g\!-\!2$, the sum of  (\ref{aIIIadmuon_ans}),
(\ref{aIIIadeet}),
(\ref{aIIIademm}),
and (\ref{aIIIademt}), is given by
\begin{equation}
       a_e^{(10)} [\text{Set III}({\rm a}_d) \text{(mass-dep)}] 
 =  0.003~28~(1),
\label{3ad_mass_dep}
\end{equation}
while the mass-dependent contributions of Set III(${\rm a}_s$)
to $a_e$ is the sum of  
(\ref{aIIIasmuon_eem}),
(\ref{aIIIasmuon_emm}),
(\ref{aIIIasmuon_eet}), and
(\ref{aIIIasmuon_emt}):
\begin{equation}
       a_e^{(10)} [\text{Set III}({\rm a}_s) \text{(mass-dep)}] 
 = 0.004~32~(10).
\label{3as_mass_dep}
\end{equation}
The total contribution of Set III(a) to $a_e$ is 
the sum of  (\ref{3adeee}),
(\ref{3aseee}),
(\ref{3ad_mass_dep}),
(\ref{3as_mass_dep}):
\begin{equation}
       a_e^{(10)} [\text{Set III(a) (all terms)}] 
 = 2.1349~(2).
\label{3a_all}
\end{equation}

Similarly, from 
(\ref{aIIIb_ans_em}),  (\ref{aIIIb_ans_et}), and  (\ref{3beee})
we obtain
\begin{equation}
       a_e^{(10)} [\text{Set III(b) (all terms)}] 
 = 3.3299~(5).
\label{3b_all}
\end{equation}

The total contribution of Set III(${\rm a}_d$) to the muon $g\!-\!2$, the sum of
 (\ref{aIIIad_mee}),
(\ref{aIIIad_mme}),
(\ref{aIIIad_met}),
(\ref{aIIIad_mmt}),
(\ref{aIIIad_mtt}),
 and (\ref{3adeee}), is
\begin{equation}
       a_\mu^{(10)} [\text{Set III}({\rm a}_d) \text{(all terms)}] 
 = 55.360~(19),
\label{mu3ad_all}
\end{equation}
while the total contribution of Set III(${\rm a}_s$) to the muon $g\!-\!2$, the sum of
(\ref{aIIIas_mee}),
(\ref{aIIIas_mem}),
(\ref{aIIIas_met}),
(\ref{aIIIas_mmt}),
(\ref{aIIIas_mtt}),
 and (\ref{3aseee}), is
\begin{equation}
       a_\mu^{(10)} [\text{Set III}({\rm a}_s) \text{(all terms)}] 
 = 57.058~(26).
\label{mu3as_all}
\end{equation}
The total contribution of Set III(a) to the muon $g\!-\!2$, the sum of
(\ref{mu3ad_all}) and  (\ref{mu3as_all}), is thus
\begin{equation}
       a_\mu^{(10)} [\text{Set III}({\rm a}) \text{(all terms)}] 
 = 112.418~(32).
\label{mu3a_all}
\end{equation}

The total contribution of Set III(b) to the muon $g\!-\!2$, from
(\ref{aIIIb_me}), (\ref{aIIIb_mt}),
and (\ref{3beee}), is
\begin{equation}
       a_\mu^{(10)} [\text{Set III}({\rm b}) \text{(all terms)}] 
 = 15.4074~(45).
\label{mu3b_all}
\end{equation}

The contribution of Set III(a) to the muon $g\!-\!2$ is very large,
which is not unexpected.  In particular, the
orders of magnitude of contributions
from  the dominant $(mee)$ terms of Set III(${\rm a}_d$) and Set III(${\rm a}_s$), 
as well as the $(me)$ term of Set III(b),
can be estimated crudely since their leading
$log (m_\mu/m_e)$ term is determined by the renormalization procedure
\cite{Kinoshita:1967,Kinoshita:2005sm}:
\begin{eqnarray}
       a_\mu^{(10)} [\text{Set III}({\rm a}_d)^{(mee)}] &\sim& 3 K_2^2 a_e^{(6)} (no~loop) \sim 34,
\nonumber  \\
       a_\mu^{(10)} [\text{Set III}({\rm a}_s)^{(mee)}] &\sim& 3 K_{2,2} a_e^{(6)} (no~loop) \sim 34,
\nonumber  \\
       a_\mu^{(10)} [\text{Set III}({\rm b})^{(me)}] &\sim& 3 K_4 a_e^{(6)} (no~loop) \sim 7 ,
\label{estimate1}
\end{eqnarray}
with
\begin{eqnarray}
       K_2 &\sim& \frac{2}{3} \ln (m_\mu/m_e) - ...,
\nonumber  \\
       K_{2,2} &\sim& K_2^2,
\nonumber  \\
       K_4 &\sim& \frac{1}{2} \ln (m_\mu/m_e) - ...,
\label{leadingfactor}
\end{eqnarray}
and \cite{Laporta:1996mq} 
\begin{equation}
    a_e^{(6)} (no~loop) = 0.904~979~....   ,
\end{equation}
where $no~loop$ means diagrams without closed lepton loops of
vacuum-polarization type.
The factor 3 accounts for the increase in the number of diagrams
caused by insertion of vacuum-polarization loops. 
As is expected from (\ref{estimate1}) and (\ref{leadingfactor}),
the values of 
(\ref{aIIIad_mee}) and (\ref{aIIIas_mee}) are
of the same order of magnitude.

\begin{acknowledgments}
This work is supported in part by the JSPS Grant-in-Aid for Scientific 
Research (C)19540322, (C)20540261, and (C)23540331.
The part of material presented by T.~K. is based on work supported
by the U.~S.~National Science Foundation under the Grant NSF-PHY-0757868,
and the International Exchange Support Grants (FY2010) of RIKEN.
T.~K. thanks RIKEN for the hospitality extended to him while
a part of this work was carried out.
Numerical calculations  are conducted 
on the RIKEN Supercombined Cluster System (RSCC)
and the RIKEN Integrated Cluster of Clusters (RICC) supercomputing systems.
\end{acknowledgments}


\bibliographystyle{apsrev}
\bibliography{b}

\end{document}